\documentclass[conference]{IEEEtran}
\IEEEoverridecommandlockouts

\usepackage{cite}
\usepackage{amsmath,amssymb,amsfonts}
\usepackage{algorithmic}
\usepackage{graphicx}
\usepackage{textcomp}
\usepackage{xcolor}

\usepackage[russian,english]{babel}
\usepackage{subcaption}
\usepackage[group-separator={,}]{siunitx}
\usepackage{ulem} 
\usepackage{float}
\usepackage{hyperref}

\def\BibTeX{{\rm B\kern-.05em{\sc i\kern-.025em b}\kern-.08em
\kern-.1667em\lower.7ex\hbox{E}\kern-.125emX}}

\begin{document}



\title{Russo-Ukrainian War: Prediction and explanation of Twitter suspension
}

\author{
\IEEEauthorblockN{\textbf{Alexander Shevtsov}}
\IEEEauthorblockA{\textit{University of Crete (CSD-UOC)} \\
\textit{Technical University of Crete (TUC) }\\
Heraklion, Greece \\
asevtsov@tuc.gr} \\

\IEEEauthorblockN{Ioannis Kontogiorgakis}
\IEEEauthorblockA{\textit{Technical University of Crete (TUC)}\\
Chania, Greece \\
}\\

\and
\IEEEauthorblockN{Despoina Antonakaki}
\IEEEauthorblockA{\textit{Technical University of Crete (TUC) } \\
\textit{ICS-FORTH}\\
Chania, Greece \\
dantonakaki@tuc.gr}
\\
\IEEEauthorblockN{Polyvios Pratikakis}
\IEEEauthorblockA{\textit{ICS-FORTH} \\
Heraklion, Greece \\
}

\and
\IEEEauthorblockN{Ioannis Lamprou}
\IEEEauthorblockA{\textit{University of Crete (CSD-UOC)} \\
\textit{Technical University of Crete (TUC) }\\
Chania, Greece \\
ilamprou1@tuc.gr
} \\
\IEEEauthorblockN{Sotiris Ioannidis}
\IEEEauthorblockA{\textit{Technical University of Crete (TUC) } \\
Chania, Greece \\
}
}

\maketitle

\begin{abstract}
On 24 February 2022, Russia invaded Ukraine, starting what is now known as the Russo-Ukrainian War, initiating an online discourse on SNs. Twitter one of the most popular SNs, with an open and democratic character, enables a transparent discussion among its large user base. Unfortunately, this often leads to Twitter's policy violations, propaganda, abusive actions, civil integrity violations, and consequently to user accounts' suspension and deletion. This study focuses on the Twitter suspension mechanism and the analysis of shared content and features leading to an accurate machine-learning suspension prediction. Toward this goal, we have obtained a dataset containing 107.7M tweets, originating from 9.8 million users, using Twitter API. We extract the categories of shared content of the suspended accounts and explain their characteristics, through the extraction of text embeddings in junction with cosine similarity clustering. Our results reveal scam campaigns taking advantage of trending topics regarding the Russia-Ukrainian conflict for Bitcoin and Ethereum fraud, spam, and advertisement campaigns. Additionally, we apply a ML methodology including a SHapley Additive explainability model to understand and explain how user accounts get suspended.
\end{abstract}


\section{Introduction}
On 24th February 2022, Russia's invasion of Ukraine, also known as the Russo-Ukrainian War, triggered extensive discussions on social media. Twitter, being one of the most prominent SNs, provides an open and democratic platform for transparent discussions among its vast user base. However, this openness can lead to policy violations, propaganda, abusive actions, and violations of civil integrity, resulting in the suspension and deletion of user accounts. This study focuses on understanding the mechanism behind Twitter suspension and analyzing the shared content and user account features that may lead to such suspensions.

To achieve this goal, we obtained a dataset comprising 107.7M tweets originating from 9.8M users, collected via Twitter API. We analyze the content shared by suspended accounts and investigate their characteristics using text embeddings and cosine similarity clustering. Our findings reveal the presence of scams and advertisement campaigns exploiting trending topics related to the Russia-Ukraine conflict towards Bitcoin and Ethereum fraud. Additionally, we employ various ML classification models and a SHapley Additive explainability \cite{lundberg2017unified} model to comprehend and explain the factors contributing to user account suspensions.

In response to the Ukraine war, Twitter has announced new moderation policies to reduce the amplification of Russian government accounts and ban certain tweets containing images of prisoners of war, that also apply to governments engaged in armed conflict restricting access to the open internet. Twitter aims to strike a balance between supporting essential documentation of events during the conflict and preventing the exploitation by state actors. State-run accounts will no longer be recommended on Twitter's home timeline, explore page, or search results if they originate from countries restricting internet access during armed conflicts.

During the Russian invasion, many researchers sharing primary material from SNs open-source intelligence (OSINT) \cite{Verge22} have faced unexpected account suspensions. Twitter later clarified that these suspensions were unintentional and not part of a coordinated campaign. The platform reinstated access to affected accounts promptly. Twitter's synthetic and manipulated media policy, which addresses the sharing of misinformation, was cited as the possible reason for the suspensions. However, it remains unclear how the violation of this policy was deemed.

Overall, our study contributes to understanding the phenomenon of Twitter suspensions by identifying spam, advertisement campaigns, and scam activities related to the Russia-Ukraine conflict. We provide insights into the characteristics of suspended accounts and analyze the content shared by these accounts. By employing ML models and explainability techniques, we shed light on the factors influencing the decision to suspend user accounts on Twitter.

Implementation source code and the utilized features are available at \href{https://github.com/alexdrk14/TwitterSuspension}{https://github.com/alexdrk14/TwitterSuspension} repository. The dataset is shared in the tabular format containing only the extracted feature values, without any personal user or tweet identifiers for privacy protection.


\section{Related Work}

Similar works focus on the suspension factors based on Twitter policy topics \cite{chowdhury2020twitter, volkova2017identifying}, including the violation of Twitter rules and policies \cite{chowdhury2021examining}, hateful and abusive activities \cite{founta2018large,davidson2017automated, nakov2021detecting}, violation of civil integrity \cite{ferrara2020characterizing} and spam campaigns \cite{chu2012detecting, zhang2012detecting, thomas2011suspended, antonakaki2016exploiting, ShevtsovICWSM22, antonakaki2021survey}. One of the initial and highly influential works was from Vern Paxson \cite{thomas2011suspended}, examining a dataset of 1.8B tweets. They characterize the spamming methodologies of 1.1M suspended accounts posting 80M tweets, analyze the features, evaluate the abuse of URLs, and provide an in-depth analysis of five spam campaigns. This study first reveals and analyses the marketplace of \textit {spam-as-a-service} on Twitter. Also, in \cite{thomas2014consequences} the authors detect compromised accounts in a dataset of 14M victims on Twitter and apply clustering for the suspended and deleted users.

Similar to our study, in \cite{volkova2017identifying} after they obtain a Russian, Spanish, and English Twitter corpus on the Russia-Ukraine conflict during 2014, they analyze suspension and deletion factors. The authors utilize multiple approaches of feature extraction based on n-grams, Latent Semantic Analysis, text embeddings, topics, emotions, and images and reveal the major differences between the deleted and suspended accounts. In \cite{zannettou2019disinformation} in 2.7K suspended Twitter accounts they study how dedicated accounts ('trolls') are involved in spreading disinformation, the type of content they disseminate, and their influence on the information ecosystem.  In \cite{im2020still} they explore Russian trolls and accounts acting on behalf of the Russian state, in a dataset of 170K control Twitter accounts,  highlighting interesting quantitative covariates among the flagged/suspended accounts as an indication of Russian trolls.

\begin{table}[tb]
\centering
\caption{Set of hashtags used in our data collection query in Russian, Ukrainian, and English. Translation for non-English HTs is in parenthesis.}
\begin{tabular}{ |p{0.95\linewidth}| } 
 \hline
\#Ukraine, \#Ukraina, \#ukraina, 
 \#\foreignlanguage{russian}{Украина}(Ukraine),
 \#\foreignlanguage{russian}{Украине}(Ukraine),
\#PrayForUkraine,
\#UkraineRussie,
\#StandWithUkraine,
\#WWIII,
\#StandWithUkraineNOW, 
\#RussiaUkraineConflict,
\#worldwar3,
\#\foreignlanguage{russian}{Война}(War),
\#RussiaUkraineCrisis,
\#StopPutin,
\#RussiaInvadedUkraine,
\#BlockPutinWallets, 
\#UkraineRussiaWar, 
\#Putin, 
\#Russia, 
\#\foreignlanguage{russian}{Россия}(Russia), 
\#StopRussianAggression, 
\#StopRussia, 
\#Ukraine\_Russia, 
\#Russian\_Ukrainian, 
\#SWIFT, 
\#NATO, 
\#FuckPutin, 
\#solidarityWithUkraine, 
\#PutinWarCriminal, 
\#PutinHitler, 
\#BoycottRussia,
\#StopWar,
\#with\_russia, 
\#FUCK\_NATO,
\#\foreignlanguage{russian}{ЯпротивВойны}(ImAgainstWar),
\#StopNazism,
\#myfriendPutin,
\#UnitedAgainstUkraine,
\#\foreignlanguage{russian}{ЯМыРоссия}(IWeRussia), 
\#\foreignlanguage{russian}{ВеликаяРоссия}(GreatRussia),
\#\foreignlanguage{russian}{Путинмойпрезидент}(PutinIsMyPresident),
\#\foreignlanguage{russian}{россиявперёд}(GreatRussia),
\#\foreignlanguage{russian}{ПутинНашПрезидент}(PutinOurPresident),
\#\foreignlanguage{russian}{ЗаПутина}(ForPutin),
\#\foreignlanguage{russian}{ПутинВведиВойска}(PutinBringTroops),
\#\foreignlanguage{russian}{СЛАВАРОССИИ}(GloryToRussia), 
\#\foreignlanguage{russian}{СЛАВАВДВ}(GloryToAirborneForces) \\ \hline
\end{tabular}
 
 \label{table:api_hashtags}
 \end{table}

An additional work analyzing the Twitter suspension mechanism was done during the 2020 US Presidential Election period \cite{chowdhury2021examining}. This work shows that suspended accounts have a higher probability of posting hateful tweets, using hashtags related to conspiracy theories belonging mostly to younger accounts, in terms of account registration date.

Additionally, explainability models in ML have also been incorporated in Twitter research towards prediction and sentiment analysis \cite{fiok2021analysis,}, towards identification of bots \cite{ShevtsovICWSM22, kouvela2020bot}, or fake news detection \cite{reis2019explainable, puraivan2021fake}. 

Recent studies concentrate on explainable ML implementations, where the models are utilized to describe the differences between suspended and normal accounts. For example, in \cite{kapoor2021ll}, they analyze the Indian 2019 general elections dataset and compare the differences between three categories of accounts; normal, suspended, and restored. They show the differences between all three classes with SHAP explainability method. This approach reveals that suspended accounts, in comparison with other user categories have a higher retweet rate proportion, average unique hashtags count, and average number of tweets per hour.

Other relevant works on Twitter suspension warnings and patterns include  \cite{yildirim2021short}, where they demonstrate how suspension warnings on Twitter can influence hate speech. In \cite{chatzakou2017mean} they study the properties of bullies and aggressors, by extracting text, user, and network-based attributes, as well as the features that distinguish them from regular users. They compare their results with the suspension and deletion of accounts of seemingly undetected users on Twitter.
Additionally, authors in \cite{wei2016exploring} explore the patterns of suspended users on Twitter, while they show that removing suspended users has no significant impact on the network structure. 
Linking Twitter and the suspended accounts, a study in a dataset of 41,352 suspended spammer accounts, \cite{ghosh2012understanding} reveals the users connected to them, investigates link farming in the Twitter network, and then explores the mechanisms to discourage this activity. Finally, in \cite{benigni2017online} they show a large Twitter community whose activity supports ISIS propaganda diffusion in varying degrees. They apply Iterative Vertex Clustering and Classification (IVCC) and leverage clustering and Twitter suspensions to infer positive case instances that give partition to the training set with 96\% accuracy. 
In the aforementioned works, they limit their methodology in terms of the extracted characteristics of user profiles, shared content, activity, and user relations. Additionally, most of the studies do not include explainability methods of implemented ML classification models in their analysis. 
In comparison with the presented approaches, we extract and evaluate a large volume of user features to identify which category is more informative in terms of account suspension. Additionally, we develop ML methods based on each of the feature categories and identify their performance over two different time periods. Finally, we exploit the most accurate models to provide explainability of the model decision with the use of the SHAP game-theoretical approach.

\begin{table}[tb]
\centering
\caption{Number of extracted and selected features for each model category.}
\begin{tabular}{ |c|c|c| } 
 \hline
 Feature category & \# Total features & \# Model selected \\ \hline
 Profile & 54 & 36 \\ \hline
 Activity timing & 139 & 109 \\ \hline
 Textual  & 67 & 53 \\ \hline
 Post embedding & 385 & 328 \\ \hline
 Graph embedding & 150 & 140 \\ \hline
 Combination & 1565 & 197 \\ \hline
\end{tabular}
 
 \label{table:all_features}
\end{table}
\section{Data}
For the purposes of our study, we collect a Twitter corpus related to the Russo-Ukrainian War. We use the Twitter streaming API to retrieve public tweets using popular hashtags related to the selected topic (Table \ref{table:api_hashtags}). Our data collection contains $\num[group-separator={,}]{107735220}$ tweets from $\num[group-separator={,}]{9851176}$ users starting from February 23, 2022. 
To label suspended accounts, we also use Twitter compliance API on a daily basis to identify the exact date of account suspension. Based on this labeling method, we identify $\num[group-separator={,}]{433466}$ suspended accounts and 86,630 deactivated accounts. To extract normal users with a high probability of being normal, we remove all suspended/deactivated accounts.

\section{Methodology}
\label{methodology}

 As mentioned earlier, in this study we focus on user-shared content, daily activity, and characteristics among the suspended accounts on Twitter, during the 2022 Russia-Ukrainian War. Towards this direction, we extract raw text from tweets of $\num[group-separator={,}]{433466}$ suspended accounts with a total volume of 4.8M documents, by applying some primitive filtering (drop URL links and user mentions).

\subsection{Shared content}
Initially, we are interested in the shared content across the suspended accounts. Due to the large volume of the extracted documents, the identification of shared content via manual analysis is a challenging task. To overcome this issue we translate posted documents via text embeddings. In our case, we use LaBSE (Language-agnostic BERT Sentence Embedding) \cite{feng2020language} model which allows us to overcome two main issues; the multi-language similarity (since our dataset contains posts from a variety of languages) and the normalization of text length and semantics similarity (embeddings of texts with similar semantics are very close to each other). A common approach to identifying similar topics is to use similarity and clustering methods. However, this methodology is highly time-consuming, especially in our case with large volumes and high vector dimensionality of the extracted embeddings. To handle this issue we apply PCA dimensionality reduction.

This procedure reduces vector dimensionality from 768 to only 20 and reduces the complexity of the text clustering. The developed vectors are clustered via the cosine similarity clustering algorithm which identifies $\num[group-separator={,}]{48486}$ unique clusters. We utilize the centroid of each cluster to analyze the text of the shared content. 

\subsection{Suspension prediction}
Besides the shared content identification, we also investigate whether the feature categories can contribute to a powerful and accurate model for detecting user suspension. This procedure includes the implementation of a model based on each separate feature category, as well as, a combination model that contains all extracted features. Our methodology consists of a model creating a pipeline including feature selection, K-Fold cross-validation, and measurement of the performance on two separated dataset portions.

To develop a generalizable model, we select a restricted time window for the feature extraction and performance estimation. The time restriction helps reduce the model's complexity. In our case, we choose a time window of 21 days, which reduces the time required for user monitoring (in the case of a real-time application) and helps the ML model capture generic 
patterns.

To identify the features that are effective in a long-term scenario, we extract two separate data portions. The first covers a 21-day monitoring period. This is because we need to simulate a real-case scenario where the initial data are collected for feature selection, model fine-tuning, and initial evaluation. The second portion is used as unseen data for proper model performance evaluation and comparison. Testing the model on the second part of 21 days of data, allows us to measure its performance in a real-case scenario, where the model is fed with data points from different periods of time containing diverse activity patterns. The first dataset includes data from February 23, 2022, until March 15, 2022, and represents the initial training sample. The second portion also contains exactly 21 days, but from a different time period, starting on March 16, 2002, until April 6, 2022. The rest of the paper will refer to these parts as \textit{portion A} and \textit{portion B}.

The extracted data contains 37,195 and 11,716 suspended accounts for \textit{portion A} and \textit{portion B} respectively. Due to the extreme imbalance between normal and suspended accounts, we utilize an under-sampling methodology and select randomly an equal volume of normal users for each data period. Based on the \textit{portion A} data, we extract the following feature categories: user profile, activity timing, textual, post embeddings, and graph embeddings.

Our main goal in this study is to identify which of the presented feature categories are more informative in terms of Twitter account suspension, as well as which feature values have the greatest impact on the model's decision. To achieve this, we create a model for each feature category and a combination model that contains all the extracted features. The following sections explain the extracted feature categories.

\subsection{Profile features}\label{profileFeatures}

In this section, we describe the feature extraction of Twitter users' profile objects. The profile objects contain user metadata, including the username, the description, the number of followers and friends, etc. The profile features provide information about the account creator, e.g. how similar the username and screen name are; and whether an account provides a long description, or uses a default profile and background picture. In the case of automatically created accounts, these features may be shared between multiple accounts, while the username and screen name have high similarity. Additionally, it is possible to evaluate the account dynamics according to the profile age, number of followers, number of friends, or number of post activities, using the following formula:

\begin{equation}
by\_age(A) = \frac{A}{Days\ since\ account\ creation}
\end{equation}

where A is the number of specific actions performed by a user (number of tweets/followers/friends etc.).

In our case, we can also calculate the growth of these parameters, during the monitoring period and calculate the growth of friends, followers, and the number of tweets between the first and the last object of the monitoring feature extraction period. The formula that calculates the growth (G) is:

\begin{equation}
G(A_{start}, A_{end}) = \frac{A_{start} + A_{end}}{A_{start}}
\end{equation}

where $A_{start}$ is the value of actions from the first user object and $A_{end}$ is the last action value, during the feature extraction period. Based on our \textit{portion A} data, we extract in total 54 unique profile features (Table \ref{table:all_features}). 

\begin{table}[tb]
\centering
\caption{Examples of detected discussion topics from suspended accounts. Sensitive metadata like account, mention, hashtag, and URL info is removed for privacy reasons.}
\begin{tabular}{|c|p{6.2cm}|}
\hline
Category & Text \\ \hline 

Crypto/NFT & Knock knock knock... Anybody is there? Your lucky door knocking ; RETWEET TAG 3 friend  1000\$ \#Bitcoin \#Airdrop \#StopWar \#UkraineRussia \#StopRussia \#Crypto \#NFTs \#NATO \#worldwar3 \#PrayingForUkraine \#Putin \#Giveaway \#ETH \#cryptocurrency \\ \hline 

Spam & How to find over 100 ways to earn money with URL via COMPANY \#Ukraine \#RussianArmy \#AssassinsCreed \#KingCharlesIII \#QueenElizabeth \#earthquake \#USOpen \#QueenElizabethII \\ \hline


Injection& \#Ukraine needs weapons and humanitarian assistance to defend against \#Putin. Russian troops shoot a nuclear power plant. Stop innocent civilian deaths. People around the world ask NATO to close the airspace over Ukraine. MENTION, provide \#SafeAirliftUkraine \\ \hline
\end{tabular}

\label{tab:examples_content}
\end{table}

\subsection{Activity timing features}\label{activityTiming}

A particular feature category is based on the statistical measurements of a user's actions (shared posts). We measure two categories of user activity patterns. The actions according to which part of the day they happen (hour) and the index of the day within a week. The second category measure is the statistics of the collected user posts (tweets, retweets, and quotes). 
In the case of activity per hour of the day, we measure the number of user actions (tweets, retweets, and quotes), according to the daily hours. Based on these measurements, we calculate the percentage of the user's activity and identify the hours within a day that users are most active. This feature provides the unique characteristics of the user activity during the day.

Similarly, we compute the user activity, according to the days of the week. This feature may also provide unique information about the user activity; for example, identify whether the users are more active during the weekends.

Furthermore, we measure the statistics of the user actions during the feature extraction period. For this purpose, we utilize a certain statistical measurement based on the action time (retweet time and quote time), where we calculate the time difference between the original post creation and the user response action time. This metric shows the average time between the user's reaction to other users' activity. The accounts that have a very "quick" reaction time are probably automated accounts that are controlled by computer software.

Additionally, we identify the user tweets, retweets, and quotes that are the most common forms of user activity, by measuring their percentage.
These described measurements provide information about the user activity routine,  like the active hours of the day, the days of the week, and the general user profile preferences of content sharing.

\begin{table}[b]
    \centering
    \caption {The most popular discussion topics according to our manual analysis, combined with the total amount of analyzed tweets and identified clusters.}
    \begin{tabular}{|c|c|c|}
    \hline
        Topic &  \# Tweets & \# Clusters\\ \hline
        Giveaway/Crypto/NFT & $\num[group-separator={,}]{165226}$ & 18 \\ \hline 
        Content injection & $\num[group-separator={,}]{71931}$ & 8\\ \hline
        SPAM/advertisment & $\num[group-separator={,}]{64699}$ & 9\\ \hline
        Military Intelligence & $\num[group-separator={,}]{59367}$ & 15 \\ \hline 
        Support for Ukraine & $\num[group-separator={,}]{34952}$ & 2 \\ \hline \hline 
        Total analyzed & $\num[group-separator={,}]{4808904}$ & $\num[group-separator={,}]{48486}$ \\ \hline
    \end{tabular}
    
    \label{tab:Sus_topics}
\end{table}

\subsection{Textual features}\label{textualFeatures}
Except from the information that can be extracted from the user profile and action pattern, we need to collect the content that users share, since Twitter users mostly communicate via content sharing. For this reason, we utilize two techniques of content analysis. Initially, we extract content metadata statistics, regarding the semantics users adopt to enrich their shared content and the text semantics via sentence embedding. In this section, we describe the first category based on the text metadata semantics.
We separate this feature category into three subcategories: tweets, retweets, and quotes. For each one, we extract separate metadata since they are created under different contexts (e.g. RTs are not written by the user herself, but already created by another user).

For each category, we retrieve the number of hashtags, URLs, and mentions. For each of these measurements, we take into account the minimum, maximum, average, and standard deviation. 
These metrics not only provide a picture of the volume but also describe the distribution of the particular feature usage. Also, since some of the hashtags are very popular, while others may be uniquely created and used only by a particular user, we compute the Term Frequency - Inverse Document Frequency (TF-IDF). The TF-IDF shows whether a particular hashtag is trending across other users or not. This information allows us to identify whether an account utilizes mostly popular entities (hashtags, user mentions). The last measurement of content metadata we compute is the user vocabulary size.

\subsection{Post embedding features}\label{post_emb}
In the previous section, we describe how we extract metadata from the user-shared objects. However, this analysis lacks text semantics. Our collected dataset contains posts in 65 different languages, which makes the extraction of text semantics very challenging. In order to eliminate this issue, we use pre-trained text embedding solutions. We utilize the LaBSE (Language-agnostic BERT Sentence Embedding) Neural Network model \cite{feng2020language} which is trained over more than 100 languages. The particular model was selected for two reasons; it provides a large space of already trained languages and does not require the alignment of input text sequences. 

Based on the selected model, we parse all user posts and export 384 features for each of them. Additionally, we store the categories of the user posts (tweets, retweets, or quotes) as one hot encoding. This particular information allows the transformation of textual information (that can't be processed by an ML model) into a numeric vector space and allows the processing by a classification model.

\begin{table}[ptb]
\centering
\caption{Graph embedding evaluation performance over different social relation graphs combined with different output dimensions. The highest performance is presented in bold text.}
\begin{tabular}{ |c|c|c|c| } 
 \hline
 Social Graph&  Output dimension & MRR & AUC\\ \hline
 Quote       & 50 & $0.65$ & $0.81$  \\ \hline
 Mention     & 50 & $0.54$ & $0.87$  \\ \hline
 Retweet     &  50 & $0.54$ & $0.84$ \\ \hline
 Multi-layer & 50 & $0.79$ & $0.96$ \\ \hline \hline
   
 Quote       & 100 & $0.67$ & $0.78$ \\ \hline
 Mention     & 100 & $0.53$ & $0.87$ \\ \hline
 Retweet     & 100 & $0.52$ & $0.84$ \\ \hline
 Multi-layer & 100 & $0.82$ & \boldmath{$0.97$}\\ \hline \hline
 Quote     & 150 & $0.66$ & $0.78$\\ \hline
 Mention   & 150 & $0.52$ & $0.86$\\ \hline
 Retweet   & 150 & $0.52$ & $0.84$\\ \hline
 Multi-layer & 150 & \boldmath{$0.84$} & \boldmath{$0.97$}\\ \hline \hline

 Quote & 200 & $0.66$ & $0.77$ \\ \hline
 Mention & 200 & $0.52$ & $0.85$\\ \hline
 Retweet & 200 & $0.52$ &  $0.84$\\ \hline
 Multi-layer & 200 & $0.54$ & $0.80$\\ \hline
 
 \hline
\end{tabular}

\label{table:graphemb}
\end{table}

\subsection{Graph embedding features}\label{graph_emb}
Twitter as a modern SN provides multiple social interactions between registered users (retweets, mentions, and quotes). In this section, we answer the research question of "Which type of social relations are important for suspension detection?". To answer this question we extract the social relations between the users in a graph form. In this social graph, the nodes represent the users, and the edges represent the relations between them. 

Graph representations cannot be used directly by an ML model for training and prediction. In order to solve this issue modern solutions utilize Neural Network (NN) implementations \cite{min2021pytorch}. In the case of a NN implementation, the model learns the graph relations between the nodes and provides a representation (also known as node embedding) in the form of a numerical vector for each relational node of the graph.
 
The provided representation allows the utilization of the graph relational knowledge as an input to the ML model. For this purpose we replicate PyTorch Big Graph Neural Network model \cite{lerer2019pytorch}, since the particular implementation allows us to feed and process very large graphs with millions of nodes, without any restriction concerning the number of social relations (also known as graph layers). The particular implementation allows us to experiment between the single relation and the multi-relational graph embeddings and keep the original structure and parameters of the GNN\cite{lerer2019pytorch}.

Twitter provides multiple user-to-user interactions, namely retweets, mentions, and quotes. In our implementation, we are interested in the identification of social interactions that provide the most important information about a user-to-user relationship. For this purpose, we extract four different social interaction graphs, quotes, mentions, retweets, and a combination of them (also known as a multi-layer graph). As a metric of social interaction importance, we keep learning the performance (MRR and AUC) of our Neural Network model for each graph, since the performance of the NN model affects immediately the performance of the ML classification model.

We permit our model to learn accurately the user-to-user relations since it achieves higher MRR and AUC scores according to the presented performance in Table \ref{table:graphemb}. Specifically, the multi-relational graph as a combination of social interactions has an embedding size equal to 150.



\section{Experimental Results}

\subsection{Shared content}

In this chapter, we present the results, based on the experiments described in the previous section. Specifically, we identify that the most popular category of suspended content is correlated to \textit{'Crypto'},\textit{ 'NFT'}, and \textit{'giveaways'}. We assume that content creators in order to trigger user attention, take advantage of trending topics, and keywords referring to the Russo-Ukrainian War, especially during the first months of the war.

Additionally, we identify similar content injection, where the users share identical content and mention celebrities (Table \ref{tab:examples_content}) to gain popularity. This kind of activity pattern is very similar to botnet activity where a particular group of accounts shares identical content at almost identical times and takes advantage of popular topics and users, as shown in similar studies \cite{shevtsov2022discovery}. Besides similar content, we also identify large spam and advertisement campaigns where accounts maliciously ride on the success of popular hashtags and trending topics to promote their products \cite{antonakaki2016exploiting}.

\begin{table*}[!htbp]
\centering
\caption{The model performance is measured during the K-Fold Cross Validation (avg. score). We show the test set of the initial \textit{portion A} data, and the second test containing \textit{portion B} data. Each model is trained/tested over the described set of features.}
\begin{tabular}{|c|c|c|c|c|c|c|c|c|c|}
\hline
Model & \multicolumn{3}{c|}{Validation}  & \multicolumn{3}{c|}{Test} & \multicolumn{3}{c|}{Second Test}\\
\cline{2-10}
 & F1 & ROC-AUC & Acc. & F1 & ROC-AUC & Acc. & F1 & ROC-AUC & Acc. \\
\hline

Profile & $\underline{0.86}$ & $\underline{0.93}$ & $\underline{0.86}$ & $\underline{0.86}$ & $\underline{0.94}$ & $\underline{0.87}$ & \boldmath{$0.79$} & \boldmath{$0.90$} & \boldmath{$0.81$} \\ \hline
 
Activity timing & $0.67$ & $0.76$ & $0.70$ & $0.68$ & $0.77$ & $0.71$ & $0.45$ & $0.62$ & $0.58$ \\ \hline

Textual & $0.71$ & $0.82$ & $0.75$ & $0.72$ & $0.82$ & $0.75$ & $0.44$ & $0.63$ & $0.59$ \\ \hline

Post embedding & $0.85$ & $0.92$ & $0.86$ & $0.83$ & $0.91$ & $0.84$ & $0.00$ & $0.73$ & $0.50$ \\ \hline

Graph embedding & $0.77$ & $0.86$ & $0.79$ & $0.77$ & $0.87$ & $0.79$ & $0.21$ & $0.50$ & $0.50$\\ \hline

Combination & \boldmath{$0.88$} & \boldmath{$0.95$} & \boldmath{$0.88$} & \boldmath{$0.88$} & \boldmath{$0.95$} & \boldmath{$0.88$} & $\underline{0.75}$ & $\underline{0.89}$ & $\underline{0.79}$ \\ \hline

\end{tabular}

\label{table:model_performance}
\end{table*}

Since we identify that the most popular discussion topics were \textit{'Crypto'} and \textit{'NFT'}, we initialize a search for related keywords like \textit{'crypto', 'NFT',} and \textit{'donation'} within the clusters. It seems that the keywords of \textit{'crypto'} and \textit{'donation'} are very popular across the suspended users' conversations \cite{urlNFTfake}. After a dynamic search and manual inspection, we reveal that multiple accounts are sharing messages similar to:'\textit{Stand with the people of Ukraine. Now accepting Bitcoin donations: BTC}' (Table \ref{tab:examples_content}). These messages trigger the attention of users to donate money in the form of Bitcoin and Ethereum with a wallet address. More specifically, we count 14 Bitcoin and 9 Ethereum wallets with a total of 16 transactions since the Russian invasion and collect in total 0.02896581 Bitcoins and 0.97419 Ethereum coins. More specifically, we expose these wallets posting tweets that support both sides, Ukraine and Russia.
In Table \ref{tab:examples_content} we show some examples of tweets in an effort to steal non-fungible tokens (NFTs) (usernames and mentions that have been removed).

Furthermore, we execute measurements of content toxicity posted by suspended accounts, in order to identify whether toxicity plays an important role in the current suspension decisions. To answer this question we use a Neural Network model \cite{Detoxify}, already trained over multiple languages. Our analysis shows that only 2.1\% of the suspended account posts are toxic. Based on this result we assume that toxicity did not play an important role in the Twitter suspension decision.

\subsection{Performance}\label{subsec:performance}

Besides the content shared of suspended accounts, we also focus on building an ML model and a feature set that can achieve high detection performance. To evaluate the performance of our model, we utilize multiple testing sets with K-Fold cross-validation.
During the initial step, we measure the performance of our models based on the average F1-score, during the K-Fold cross-validation (K=5). The cross-validation measurement provides the first impression of the model performance over the training and validation of the data set portion. In the following step, we measure the performance of each model, based on the testing (hold-out) data not seeded during the feature selection, the model fine-tuning, and the cross-validation. The performance of the test data provides crucial information about the model generalization ability and the over/under-fitting issues.
Based on these results (see Table \ref{table:model_performance}) of the K-Fold cross-validation and test data-set performance, we identify that the best performance is achieved by a model that contains the combination of the features and the user profile model, with a slightly better performance of the combination model over the validation and test data-sets.

Since both validation and test data portions were extracted from the exact same time period, we are interested in the identification of the performance of new data created by the users in different time periods. For this reason, we retrieve a second testing dataset (\textit{portion B}) containing unseen normal and suspended users from the next 21 days. The extracted dataset contains exactly the same feature categories with \textit{portion A}. Each feature is extracted based only on the \textit{portion B} 21 days period, in order to avoid information leaking, except the graph embedding. In the case of graph embeddings, we keep the training data user relation as well, in order to properly allocate the node positions in an N-dimensional output space.

The extracted dataset contains exactly the same feature space and keeps class balance within the dataset, with the usage of an under-sampling method, as well as within the training, validation, and test data portions. During the performance measurement, we train our model over the entire \textit{portion A} dataset and measure how the model performs on unknown data originating from a different time period (\textit{portion B} data). Based on the results presented in Table \ref{table:model_performance}, we notice that the performance of multiple models drops significantly, especially in the case of embeddings (text and graph). The drop in the embedding features indicates that they are strongly correlated to the user content (or activity, in the case of graph embeddings). Additionally, we notice that a slight change in the user actions leads to completely different results of the embedding extraction technique. 

Despite this, some models achieve decent performance, even in the next time period, such as the profile and the combination model. The profile model provides the lowest performance drop, in comparison with any other model. This outcome allows us to assume that the user profiles remain almost the same even after a 21-day period. 

\begin{figure}[tp]
\centering
\includegraphics[width=3.0in]{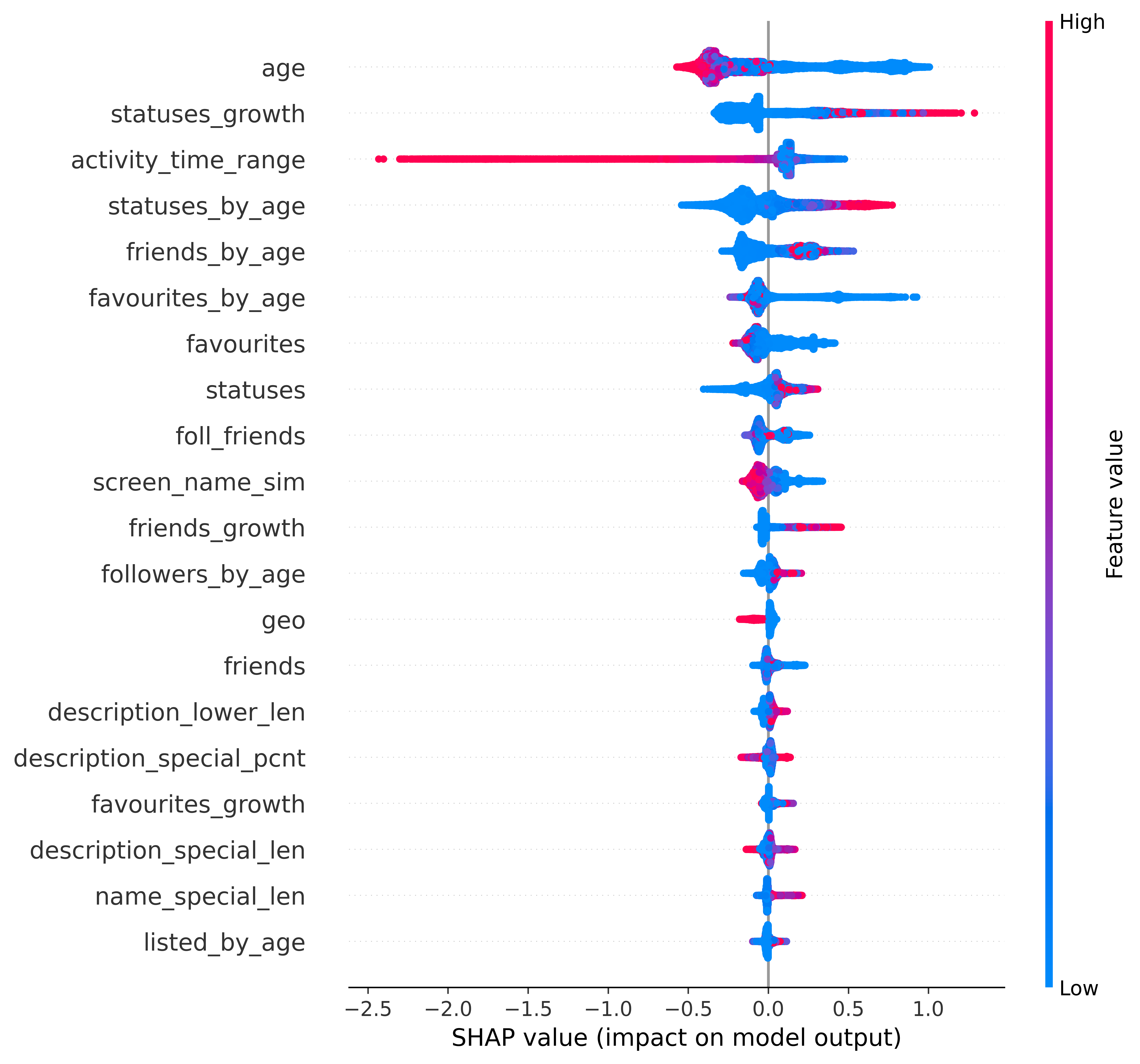}
\caption{Explainability figure (SHAP) of profile feature trained models.}
\label{fig:shapProfile}
\end{figure}

\subsection{Model explainability}\label{subsec:modelExpl}

Based on the measured performance, seen in Table \ref{table:model_performance}, we manage to identify two models that allow us to detect suspended accounts. Additionally, within this study, we are interested in the identification of differences between normal and suspended accounts on Twitter. 
Toward this goal, we build an ML model to spot the differences between the feature values of suspended and normal accounts. We achieve this with the usage of state-of-the-art SHAP (SHapley Additive exPlanations) \cite{lundberg2017unified} method, based on a game theoretic approach to explain the output of the ML model, and considered one of the best models in interpretability \cite{mosca2022shap}. We select the profiles and the combination of the feature models, in order to explain these value differences.

\begin{figure}[tp]
\centering
\includegraphics[width=3.0in]{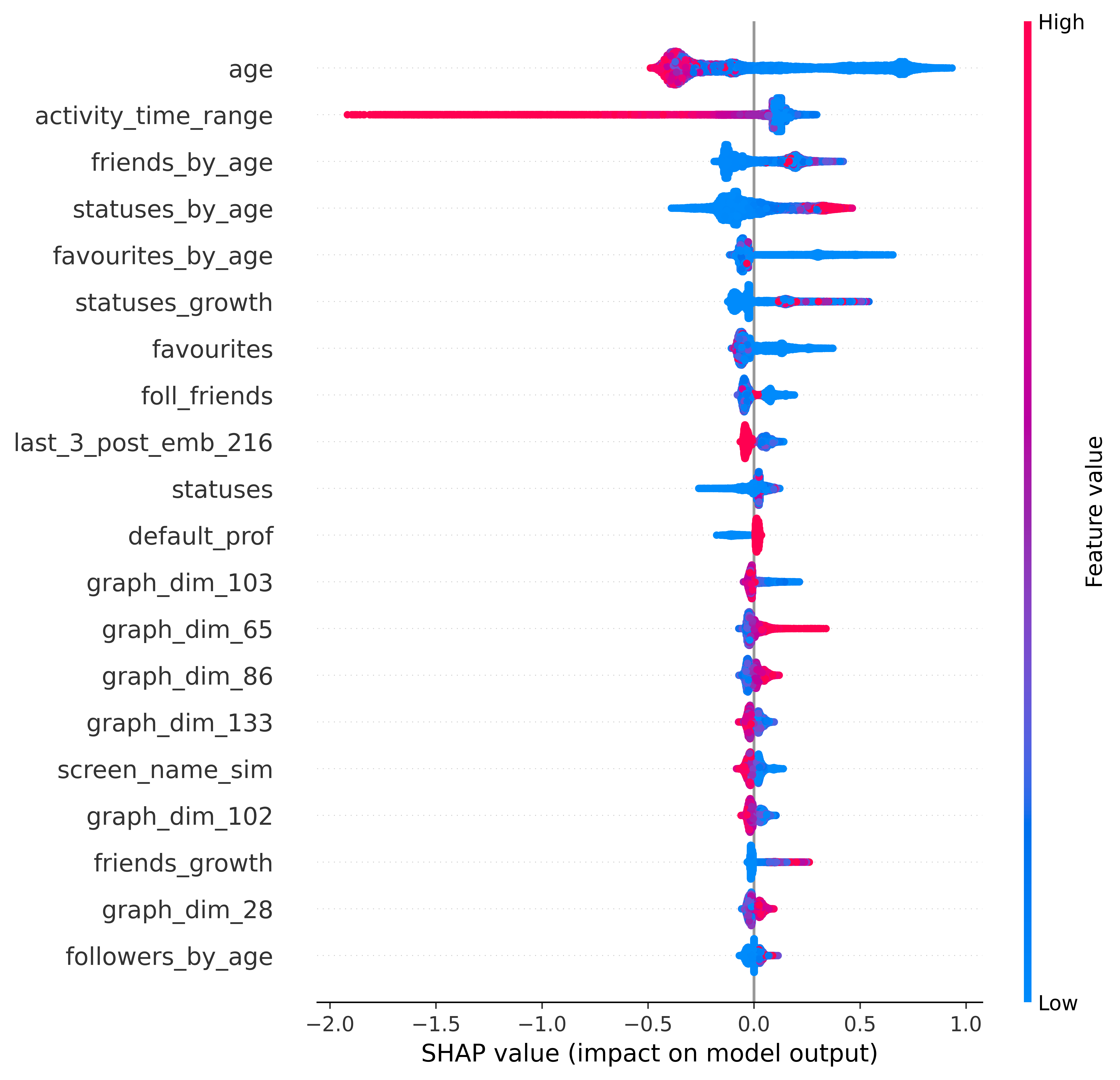}
\caption{Explainability figure (SHAP) of combination feature trained models.} 
\label{fig:shapComb}
\end{figure}

In Fig. \ref{fig:shapComb} and \ref{fig:shapProfile} we can see that both models identify similar patterns; suspended accounts have a low account age and a short period of activity (specifically notice the 'activity\_time\_range' in Fig. \ref{fig:shapComb} and \ref{fig:shapProfile}) and at the same time high values of status, according to the account age. These results show that suspended accounts are mostly fresh registered accounts posting high volumes of tweets, within a short time period. Additionally, suspended accounts grow their friends' relationship networks faster than normal users. Based on the identified results, the activity patterns of the suspended accounts on Twitter are very similar to those of the automated bot accounts. 

\section{Conclusion}
In this study, we extensively examine and extract a vast amount of characteristics for the suspended accounts involved in the online discourse during the Russo-Ukrainian War, initiated in 2022. For this purpose, we collect a large volume of user profiles along with their shared content. Based on the examined features we develop and evaluate multiple ML models and identify their user characteristics. Our methodology allows the accurate prediction of the suspended accounts by Twitter. Specifically, we utilize a state-of-the-art model decision explainability SHAP methodology in order to describe the actual model decisions based on the feature values. Our study shows that the most important characteristics behind suspension are the age of the account combined with the user actions (posts, likes, retweets, number of friends, and followers). Furthermore, we show that the most informative feature category is the combination of multiple features with 0.88 F1 and 0.95 ROC-AUC scores during the initial \textit{portion A} data. 
In the case of the \textit{portion B} data, we estimate the performance prediction on long-term scenarios, where profile features achieve the highest performance of 0.79 F1 and 0.9 ROC-AUC. Based on the achieved performance we are interested in further analysis over the larger time periods. Such an approach will allow us to evaluate and compare all of the models based on the described feature categories.

\section{Acknowledgement}


This document is the result of the research projects AI4HEALTHSEC (grant agreement ID 883273), MARVEL (grant agreement ID 957337) and GREEN.DAT.AI (grant agreement ID 101070416) co-funded by the European Commission, with (EUROPEAN COMMISSION Directorate-General Communications Networks, Content and Technology).

\bibliographystyle{IEEEtran}  
\bibliography{paper}

\end{document}